\documentclass[twoside,twocolumn,9pt]{article}
\usepackage{extsizes}
\usepackage[super,sort&compress,comma]{natbib} 
\usepackage[version=3]{mhchem}
\usepackage[left=1.5cm, right=1.5cm, top=1.785cm, bottom=2.0cm]{geometry}
\usepackage{balance}
\usepackage{mathptmx}
\usepackage{sectsty}
\usepackage{graphicx} 
\usepackage{lastpage}
\usepackage[format=plain,justification=justified,singlelinecheck=false,font={stretch=1.125,small,sf},labelfont=bf,labelsep=space]{caption}
\usepackage{float}
\usepackage{fancyhdr}
\usepackage{fnpos}
\usepackage[english]{babel}
\addto{\captionsenglish}{%
  
}
\usepackage{array}
\usepackage{droidsans}
\usepackage{charter}
\usepackage[T1]{fontenc}
\usepackage[usenames,dvipsnames]{xcolor}
\usepackage{setspace}
\usepackage[compact]{titlesec}
\usepackage{hyperref}

\usepackage{epstopdf}

\definecolor{cream}{RGB}{222,217,201}

\begin{document}

\pagestyle{fancy}
\thispagestyle{plain}
\fancypagestyle{plain}{
\renewcommand{\headrulewidth}{0pt}
}

\makeFNbottom
\makeatletter
\renewcommand\LARGE{\@setfontsize\LARGE{15pt}{17}}
\renewcommand\Large{\@setfontsize\Large{12pt}{14}}
\renewcommand\large{\@setfontsize\large{10pt}{12}}
\renewcommand\footnotesize{\@setfontsize\footnotesize{7pt}{10}}
\renewcommand\scriptsize{\@setfontsize\scriptsize{7pt}{7}}
\makeatother

\renewcommand{\thefootnote}{\fnsymbol{footnote}}
\renewcommand\footnoterule{\vspace*{1pt}%
\color{cream}\hrule width 3.5in height 0.4pt \color{black} \vspace*{5pt}} 
\setcounter{secnumdepth}{5}

\makeatletter 
\renewcommand\@biblabel[1]{#1}            
\renewcommand\@makefntext[1]%
{\noindent\makebox[0pt][r]{\@thefnmark\,}#1}
\makeatother 
\renewcommand{\figurename}{\small{Fig.}~}
\sectionfont{\sffamily\Large}
\subsectionfont{\normalsize}
\subsubsectionfont{\bf}
\setstretch{1.125} 
\setlength{\skip\footins}{0.8cm}
\setlength{\footnotesep}{0.25cm}
\setlength{\jot}{10pt}
\titlespacing*{\section}{0pt}{4pt}{4pt}
\titlespacing*{\subsection}{0pt}{15pt}{1pt}

\fancyfoot{}
\fancyfoot[LO,RE]{\vspace{-7.1pt}\includegraphics[height=9pt]{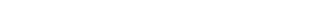}}
\fancyfoot[CO]{\vspace{-7.1pt}\hspace{13.2cm}\includegraphics{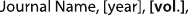}}
\fancyfoot[CE]{\vspace{-7.2pt}\hspace{-14.2cm}\includegraphics{head_foot/RF}}
\fancyfoot[RO]{\footnotesize{\sffamily{1--\pageref{LastPage} ~\textbar  \hspace{2pt}\thepage}}}
\fancyfoot[LE]{\footnotesize{\sffamily{\thepage~\textbar\hspace{3.45cm} 1--\pageref{LastPage}}}}
\fancyhead{}
\renewcommand{\headrulewidth}{0pt} 
\renewcommand{\footrulewidth}{0pt}
\setlength{\arrayrulewidth}{1pt}
\setlength{\columnsep}{6.5mm}
\setlength\bibsep{1pt}

\makeatletter 
\newlength{\figrulesep} 
\setlength{\figrulesep}{0.5\textfloatsep} 

\newcommand{\topfigrule}{\vspace*{-1pt}%
\noindent{\color{cream}\rule[-\figrulesep]{\columnwidth}{1.5pt}} }

\newcommand{\botfigrule}{\vspace*{-2pt}%
\noindent{\color{cream}\rule[\figrulesep]{\columnwidth}{1.5pt}} }

\newcommand{\dblfigrule}{\vspace*{-1pt}%
\noindent{\color{cream}\rule[-\figrulesep]{\textwidth}{1.5pt}} }

\makeatother

\twocolumn[
  \begin{@twocolumnfalse}
\vspace{1em}
\sffamily
\begin{tabular}{m{4.5cm} p{13.5cm} }

\includegraphics{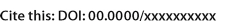}& 
\noindent\LARGE{\textbf{Equation-of-motion Coupled-cluster singles, doubles and (full) triples for doubly ionized and two-electron-attached states: A Computational implementation }} \\
 & \vspace{0.3cm} \\

 & \noindent\large{Manisha, \textit{$^{a}$} Prashant Uday Manohar\textit{$^{a}$}} and Anna.I.Krylov,\textit{$^{b\ddag}$} \\

\includegraphics{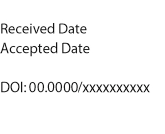} & \\

\end{tabular}

 \end{@twocolumnfalse} \vspace{0.6cm}

  ]

\renewcommand*\rmdefault{bch}\normalfont\upshape
\rmfamily
\section*{}
\vspace{-1cm}


\footnotetext{\textit{$^{a}$~ Department of Chemistry, Birla Institute of Technology \& Science-Pilani, Pilani, Rajasthan 333031, India; E-mail: pumanohar@pilani.bits-pilani.ac.in}}
\footnotetext{\textit{$^{b}$~ Department of Chemistry, University of Southern California, Los Angeles, CA 90089-0482, USA, E-mail: krylov@usc.edu}}





 We present our computational implementation of the equation-of-motion (EOM) coupled-cluster (CC) singles, doubles, and triples (SDT) method for computing doubly ionized (DIP) and two-electron attached (DEA) states within {\sl Q-CHEM}\cite{epifanovsky2021software}. These variants have been implemented within both the (conventional) double precision (DP) and the single precision (SP) algorithms and will be available in the upcoming major release of {\sl Q-CHEM}. We present here the programmable expressions and some pilot application of $CH_2$  for DIP and DEA EOM-CCSDT.


\rmfamily 

\section{Introduction}
In the past few decades, the EOM-CC 
\cite{vcivzek1966correlation,rowe1968equations,sinanouglu1970three,kummel2003biography,manohar2007dipole,manohar2010some,manohar2024spin}
suite of methods have gained popularity due to their ability to accurately study the open-shell and excited states, offering an adaptable and efficient alternative to the multi-reference (MR) CC 
\cite{mukherjee1977applications,lindgren1978coupled,pal1987multireference,pal1988molecular,jeziorski1981coupled,lindgren1987connectivity,malrieu1985intermediate,mukherjee1989use,KUMAR2019234}
approaches. Amongst them, the so-called EOM-CCSD variants (SD for singles and doubles) are very common now-a-days, and are obtained when the CC vectors as well as the EOM eigenvectors are truncated so as
to include the one-body and two-body amplitudes and provide size-intensive excitation/ionization/electron-attachment energies in, respectively the electronic excitation (EE), ionization potential (IP), electron-attachment (EA) variants. In these variants, the CC Hamiltonian spans over the singly and doubly occupied configurations in respectively, the excited, ionized and electron-attached subspaces. The eigenvalues of the CC Hamiltonian would give the energies of the respective target states.
If electronic excitation is associated with flipping of one (and only one) electron, then it leads to spin-flip SF variant of EOM-CC and provides an efficient variant to
target excited states of different multiplicities starting from a high-spin reference
wavefunction.
On the similar lines, double ionization potential (DIP)\cite{nooijen1997new,nooijen2002state,sattelmeyer2003use, musial2011multireference, mmusial2011multireference, kus2011using, kus2012perturbative} and double/two-electron-attachment (DEA)\cite{musial2011multireference,shen2013doubly,perera2016singlet,nooijen1995equation} variants are obtained when the CC Hamiltonian is formed in the subspace of doubly ionized/electron-attached configurations.

The EOM-CCSD provides and accuracy of $0.1-0.3 eV$. This means that for systems with strongly
degenerate electronic states (energy gaps of $0.3$ or less), EOM-CCSD may not be adequate and one needs to incorporate higher-body corrections at both CC as well as EOM levels. The perturbative approaches such as CCSD(T)\cite{raghavachari1989fifth} improve the total energies and energy derivatives, however, the uneven amount of
corrections in reference and target states often leads to poorer description of
energy-gaps. A balanced improvement in EOM-CCSD, is thus, only possible when
full triples are included at CC as well as EOM level resulting in the
EOM-CCSDT method.

In this article, we focus our discussion to DIP and DEA variants of EOM-CCSDT. In these methods, the occupancy of the target state configurations is different from
that of the reference configuration by $2$. As a result, the smallest EOM vector
is actually, the two-electron operator. Thus, while we have single, double and
triple substitution at CC level, the EOM level has a double, triple and quadruple
substitution. Thus, rigorously speaking, the DIP/DEA-EOM-CCSDT may actually be
named as DIP/DEA-EOM-CC(3,4). We have implemented these methods both within
the (conventional) double-precision (DP) as well as within the single-precision (SP)\cite{pokhilko2018double,30711,5976968}
algorithms. SP reduces both the storage and memory requirements in CC as well as
EOM parts of the computations almost by $50 \%$, thereby widening the applicability
of the methods to larger systems. The reduction in accuracy due to use of SP
instead of DP is highly insignificant ($O(10^{-4}) eV$).

We present very brief theory of EOM-CCSDT in the context of DIP and DEA variants in Section 2. Using methylene diradical as the model system, we discuss the excitation
energies of various target states computed using our methods and some other
available methods, in Section 3, followed by conclusions.

\section{Theory}
The DEA and DIP EOM-CC methods have been thoroughly explored in existing research. In this work, we aim to present a simplified and clear formulation of our DEA and DIP-EOM(3,4) methods for better understanding.
The m-state wavefunction $\psi_m$ is obtained by operating the $R(m)$ operator on the reference wavefunction.
\begin{eqnarray}
|\Psi_m\rangle&=& \Hat{R}|\Psi_0\rangle\\
|\Psi_m\rangle&=& \Hat{R}e^{\Hat{T}}|\Phi_0\rangle \nonumber
\label{equation1EOM}
\end{eqnarray}
where the  $|\Psi_0\rangle$  is the ground state coupled-cluster wavefunction and R(m) is the linear operator which operates on ground state CC wavefunction. In case of DIP $R_m$ operator remove two electron from the CC ground state whereas for DEA  $R_m$ operator add two electrons to ground state.
\begin{eqnarray}
    |\Psi_0\rangle&=& e^{\Hat{T}}|\Phi_0\rangle
    \label{equation2CC}
\end{eqnarray}
Now substituting the equation  \ref{equation2CC} to the schrodinger wave equation
\begin{eqnarray}
e^{-\hat{T}}\Hat{H}e^{\hat{T}}|\Phi_0\rangle &=&
(\Hat{H}e^{\hat{T}})_{C}|\Phi_0\rangle
=\Bar{H} |\Phi_0\rangle = E_{0} |\Phi_0\rangle \label{eq:hbareigen} \\
E_{CC}&=&E_{0} = \langle\Phi_{0}|\Bar{H}|\Phi_0\rangle \label{eq:ccenergy} \\
0&=&\langle\Phi^{*}|\Bar{H}|\Phi_0 \rangle \label{eq:ccteqns}
\end{eqnarray}
where we are  truncating the $\hat{T}$ upto triples, $(\hat{T}=\hat{T}_{1}+\hat{T}_{2}+\hat{T}_{3})$. For DEA and DIP state the  R(m) operator is the sum of one body, two body and three body DEA or DIP operator.
  \begin{eqnarray}
\hat{R}_m &=& \hat{R}_{2}(m) + \hat{R}_{3}(m) + \hat{R}_{4}(m), \\
\hat{R}_m(DEA) &=& \frac{1}{2} \sum\limits_{ab} r^{ab}(m) a^\dagger b^\dagger \nonumber \\
& & + \frac{1}{6} \sum\limits_{abc} \sum\limits_{i} r^{abc}_i (m)a^\dagger b^\dagger c^\dagger i \nonumber \\
& & + \frac{1}{48} \sum\limits_{abcd} \sum\limits_{ij} r^{abcd}_{ij} (m)a^\dagger b^\dagger c^\dagger d^\dagger j i.\\
\hat{R}_m(DIP) &=& \frac{1}{2} \sum\limits_{ij} r_{ij}(m) ij \nonumber \\
 & & + \frac{1}{6} \sum\limits_{ijk} \sum\limits_{a} r_{ijk}^{\phantom{...}{a}} (m)a^\dagger kji \nonumber \\
& & + \frac{1}{48} \sum\limits_{ijkl} \sum\limits_{ab} r^{\phantom{...}{ab}}_{ijkl} (m)a^\dagger b^\dagger l k j i.
\label{equation8DIP}
 \end{eqnarray}
where $i,j,k,l,...$ belongs to the occupied orbitals whereas $a,b,c,d,...$ to the virtual orbitals.
substituting the equation \ref{equation1EOM} to Schr\"{o}dinger wave equation:
\begin{eqnarray}
\hat{H}|\Psi_m\rangle &=& H \hat{R} e^{\hat{T}} |\Phi_0\rangle =\hat{H} e^{\hat{T}} \hat{R} |\Phi_0\rangle
= E_m \hat{R} e^{\hat{T}} |\Phi_0\rangle. \\ 
\bar{H} \hat{R} |\Phi_0\rangle &=& E_m \hat{R} |\Phi_0\rangle \nonumber\\
 \Bar{H} &=& e^{-\hat{T}}\Hat{H}e^{\hat{T}}  
\end{eqnarray}
where $\Bar{H}$ is the similarity transformed hamiltonian.
Action of $\hat{R}$ on $\bar{H}$ in equation \ref{equation8DIP}
 would give
\begin{eqnarray}
\hat{R} \Bar{H}  \Phi_{0} \rangle &=& E_{CC} \hat{R}  |\Phi_0\rangle 
\label{eq:RH}
\end{eqnarray}
The above equation can be transformed into matrix eigenvalue equation.
\begin{eqnarray}
  \begin{pmatrix}
\bar H_{DD} - E_{CC} & \bar H_{DT} &\bar H_{DQ} \\
\bar H_{TD} & \bar H_{TT} - E_{CC}   & \bar H_{TQ} \\
 \bar H_{QD}&\bar H_{QT}& \bar H_{QQ} - E_{CC}
 \end{pmatrix}   
  \begin{pmatrix}
  R_2\\
  R_3 \\
  R_4
   \end{pmatrix} =\omega \begin{pmatrix}
  R_2\\
  R_3 \\
  R_4
   \end{pmatrix} && \label{eq:sthmat}
   \end{eqnarray}
   where, the roots, $\omega= E-E_{CC}$, would give the DEA and DIP energies of the respective target states. Left eigenvectors are required only when calculating gradients or properties. Fully diagonalizing the CC Hamiltonian matrix in equation \ref{eq:sthmat}  is computationally unfeasible, as the focus is typically on a limited number of electronic states. Davidson's iterative diagonalization \cite{davidson197514,hirao1982generalization} method provides an efficient alternative, avoiding the need for full diagonalization while enabling the targeted computation of specific roots of the matrix. The corresponding programmable expressions within the (DEA,DIP) EOM-CCSDT framework are detailed in the Appendix.

 \section{Results and Discussion}
 \subsection*{Vertical excitation energies of methylene diradical}
 All the computations have been performed using the development version of {\sl Q-CHEM} with core orbitals frozen in the post-Hartree-Fock methods. The FCI/TZ2P
 \cite{sherrill1998structures}
 optimized ground state geometry of methylene diradical was used. The Table \ref{table:ch2vert} summarizes the results. The results obtained by SP
 algorithm are labeled as (SP) in the table and differ from the corresponding DP
 values by $10^{-4} eV$ or less.
   \begin{table}
\caption{Total energies of ground state ($\tilde{X}^{3}B_{1}$) and
Total excitation energies (eV) of $CH_{2}$}
\label{table:ch2vert}
\begin{tabular}{lccccc} \hline \hline
Method  & $\tilde{X}^{3}B_{1}$&$\tilde{a}^{1}A_{1}$ & $\tilde{b}^{1}B_{1}$ & $\tilde{c}^{1}A_{1}$ \\
\\
\hline
 
SF-EOM-CCSD   &-39.03913 &1.077&1.694&3.520\\
SF-EOM-CCSDT  &-39.04159 &1.072&1.685&3.495\\       
SF-EOM-CCSDT(SP) &-39.04159 &1.072&1.685&3.495\\    
DEA-EOM-CCSD     &-39.02832&1.004&1.613&3.391\\           
DEA-EOM-CCSDT    &-39.04170&1.073&1.682&3.494\\     
DEA-EOM-CCSDT(SP)&-39.04171&1.073&1.682&3.494\\      
DIP-EOM-CCSD      &-39.02494&1.045&1.640&3.384\\        
DIP-EOM-CCSDT     &-39.04145&1.070&1.685&3.495\\        
DIP-EOM-CCSDT(SP)   &-39.04147&1.070&1.685&3.495 \\       
\hline \hline
\end{tabular}
\end{table}
The DIP/DEA computed excitation energies (EEs) differ from the SF computed EEs by $0.0025 eV$ or less at CCSDT level using both SP and DP algorithms. This is very small compared to the differences at the CCSD level. At CCSD level, the absolute difference between the EEs computed by DEA versus SF methods is as large as $0.13 eV$, whereas for DIP versus SF, it is $0.14 eV$. The DEA versus DIP max absolute error in EEs is reduced from $0.04 eV$ at CCSD level to $0.0035 eV$ at CCSDT level.

\section{Conclusion}
We have presented our implementation of DIP-EOM-CCSDT (or DIP-EOM-CC(3,4)) and DEA-EOM-CCSDT (or DEA-EOM-CC(3,4)) methods within {\sl Q-CHEM}. Inclusion of full corrections from six-index quantities (triples at CC level and quadruples at DIP/DEA-EOM level)
results in increased agreement in the energy-gaps between same set of states computed by
the two methods by an order of magnitude, thereby reducing the maximum absolute error
from $0.004 eV$ at CCSD level to $0.0035 eV$ at CCSDT level. Errors due to use of SP algorithm versus DP seldom affects the accuracy of computed energy gaps.






\scriptsize{
\bibliography{EOMDEADIP} 
\bibliographystyle{EOMDEADIP} } 

\section*{Appendix}
The programmable expressions of DEA-EOM-CCSDT within the Davidson's iterative diagonalization algorithm are presented below:

\begin{eqnarray} 
{\sigma}^{ab} &=& ([\bar{H}_{SS}-E_{CC}]R_1)^{ab} + (\bar{H}_{SD}R_2)^{ab} + (\bar{H}_{ST}R_3)^{ab} \nonumber \\
& & \nonumber \\
 {\sigma}^{abc}_{i} &=& (\bar{H}_{DS}R_1)^{abc}_{i} + ([\bar{H}_{DD}-E_{CC}]R_2)^{abc}_{i} + (\bar{H}_{DT}R_3)^{abc}_{i} \nonumber\\
& & \nonumber \\
 {\sigma}^{abcd}_{ij} &=& (\bar{H}_{TS}R_1)^{abcd}_{ij} + (\bar{H}_{TD}R_2)^{abcd}_{ij} + ([\bar{H}_{TT}-E_{CC}]R_3)^{abcd}_{ij} \nonumber
\end{eqnarray}

For computational convenience, we express the ${\sigma}^{ab}$, ${\sigma}^{abc}_{i}$, and ${\sigma}^{abcd}_{ij}$ as follows:

\begin{align*}
{\sigma}^{ab} &= \langle\Phi^{ab}|\bar{H}|R_1\Phi_{0}\rangle + \langle\Phi^{ab}|\bar{H}|R_2\Phi_{0}\rangle + \langle\Phi^{ab}|\bar{H}|R_3\Phi_{0}\rangle \\
&= P(ab)\sum\limits_c F_{bc} r_{ac} - \frac{1}{2} P(ab) \sum\limits_{icd} I^7_{ibcd} r_i^{acd} + \frac{1}{2} \sum\limits_{cd} I^5_{abcd} r_{cd} \\
&\quad - \sum\limits_{ic} F_{ic} r_i^{bac} + \frac{1}{4} \sum\limits_{ijcd} \langle ij || dc \rangle r_{ij}^{dcab}
\end{align*}

\begin{align*}
{\sigma}^{abc}_{i} &= \langle\Phi_{i}^{abc}|\bar{H}|R_1\Phi_{0}\rangle + \langle\Phi_{i}^{abc}|\bar{H}|R_2\Phi_{0}\rangle + \langle\Phi_{i}^{abc}|\bar{H}|R_3\Phi_{0}\rangle \\
&= P(c/ab) \left[ \sum\limits_d F_{cd} r_i^{abd} - \sum\limits_{jd} I^1_{idjc} r_j^{abd} + \frac{1}{2} \sum\limits_{de} I^5_{abde} r_i^{dec} \right. \\
&\quad - \sum\limits_j H^{8'}_{jc} t_{ij}^{ab} + \sum\limits_d I^3_{idab} r_{dc} ] \\
&\quad - \frac{1}{2} \sum\limits_{nme} I^6_{mnie} r_{nm}^{eabc} + \frac{1}{2} P(a/bc) \sum\limits_{mef} I^7_{mafe} r_{mi}^{febc} \\
&\quad + \sum\limits_{ke} F_{ke} r_{ki}^{eabc} + \frac{1}{2} \sum\limits_{jk} I^{4r}_{jk} t_{ijk}^{abc} - \sum\limits_j F_{ij} r_j^{abc}
\end{align*}

\begin{align*}
{\sigma}^{abcd}_{ij} &= \langle\Phi_{ij}^{abcd}|\bar{H}|R_1\Phi_{0}\rangle + \langle\Phi_{ij}^{abcd}|\bar{H}|R_2\Phi_{0}\rangle + \langle\Phi_{ij}^{abcd}|\bar{H}|R_3\Phi_{0}\rangle \\
&= -P(ij) \sum\limits_k F_{jk} r_{ik}^{abcd} + P(a/bcd) \sum\limits_e F_{ae} r_{ij}^{ebcd} \\
&\quad - P(a/bcd) \sum\limits_m I^2_{jima} r_m^{bcd} + P(ij) P(ab|cd) \sum\limits_e I^3_{ieab} r_j^{ecd} \\
&\quad - P(ij) P(b/acd) \sum\limits_{me} I^1_{iemb} r_{mj}^{aecd} + \frac{1}{2} \sum\limits_{lk} I^4_{ijlk} r_{lk}^{abcd} \\
&\quad + \frac{1}{2} P(ab/cd) \sum\limits_{ef} I^5_{cdef} r_{ij}^{abef} + P(a/bcd) \sum\limits_e H^5_{ebcd} t_{ij}^{ae} \\
&\quad - P(ij) P(ab/cd) \sum\limits_m H^{5'}_{jmcd} t_{im}^{ab} + \frac{1}{2} P(ij) P(d/abc) \sum\limits_{nl} H^7_{jnld} t_{inl}^{abc} \\
&\quad - P(ad/bc) \sum\limits_{el} H^6_{ebcl} t_{ijl}^{aed} - P(d/abc) \sum\limits_k H^8_{kd} t_{ijk}^{abc}
\end{align*}

Intermediates used for DEA:

\begin{align*}
I^{4r}_{ij} &= \frac{1}{2} \sum\limits_{cb} \langle ij || cb \rangle r_{cb} \\
H^{8'}_{ia} &= \frac{1}{2}\sum\limits_{bc} I^7_{iabc} r_{bc} + \frac{1}{2}\sum\limits_{jbc} \langle ji || bc \rangle r_j^{abc} \\
H^8_{ia} &= \frac{1}{2}\sum\limits_{bc} I^7_{iabc} r_{bc} + \frac{1}{2} \sum\limits_{jbc} \langle ji || bc \rangle r_j^{abc} + \sum\limits_b F_{ib} r_{ba} \\
H^5_{edcb} &= \sum\limits_{mf} I^7_{mdfe} r_m^{fcb} - \sum\limits_f I^5_{dcef} r_{bf} - \frac{1}{2} \sum\limits_{mnf} \langle nm || fe \rangle r_{nm}^{fdcb} \\
H^{5'}_{kmba} &= P(ab) \sum\limits_e I^1_{kemb} r_{ea} + \sum\limits_{ne} I^6_{mnke} r_n^{eba} - \frac{1}{2} P(ab) \sum\limits_{ef} I^7_{mafe} r_k^{feb} \\
&\quad + \frac{1}{2} \sum\limits_{nef} \langle nm || fe \rangle r_{nk}^{feba} + \frac{1}{2} \sum\limits_{nfe} \langle nm || fe \rangle r_{fe} t_{nk}^{ba} \\
H^6_{bacm} &= -P(ac) \sum\limits_d I^7_{madb} r_{dc} - \sum\limits_{ne} \langle nm || be \rangle r_n^{ace} \\
H^7_{ijkb} &= \sum\limits_d I^6_{jkid} r_{db} + \frac{1}{2} \sum\limits_{da} \langle ik || da \rangle r_j^{dab}
\end{align*}

The programmable expressions of DIP-EOM-CCSDT within the Davidson's iterative diagonalization algorithm are presented below:

\begin{eqnarray} 
{\sigma}_{ij} &=& ([\bar{H}_{SS}-E_{CC}]R_1)_{ij} + (\bar{H}_{SD}R_2)_{ij} + (\bar{H}_{ST}R_3)_{ij} \nonumber \\
& & \nonumber \\
 {\sigma}^{\phantom{...}{a}}_{ijk} &=& (\bar{H}_{DS}R_1)^{\phantom{...}{a}}_{ijk} + ([\bar{H}_{DD}-E_{CC}]R_2)^{\phantom{...}{a}}_{ijk} + (\bar{H}_{DT}R_3)^{\phantom{...}{a}}_{ijk} \nonumber\\
& & \nonumber \\
 {\sigma}^{\phantom{...}{ab}}_{ijkl}  &=& (\bar{H}_{TS}R_1)^{\phantom{...}{a}}_{ijk} + (\bar{H}_{TD}R_2)^{\phantom{...}{ab}}_{ijkl} + ([\bar{H}_{TT}-E_{CC}]R_3)^{\phantom{...}{ab}}_{ijkl} \nonumber
\end{eqnarray}

For computational convenience, we express the ${\sigma}^{ij}$, ${\sigma}_{ijk}^{\phantom{...}{a}}$, and ${\sigma}_{ijkl}^{\phantom{...}{ab}}$ as follows:

\begin{align*}
{\sigma}_{ij} &= \langle\Phi_{ij}|\bar{H}|R_1\Phi_{0}\rangle + \langle\Phi_{ij}|\bar{H}|R_2\Phi_{0}\rangle + \langle\Phi_{ij}|\bar{H}|R_3\Phi_{0}\rangle \\
&= P(ij) \sum\limits_k F_{ik} r_{jk} + \frac{1}{2} P(ij) \sum\limits_{kla} I^6_{klia} r^{\phantom{...}{a}}_{klj} + \frac{1}{2} \sum\limits_{kl} I^4_{ijkl} r_{kl} \\
&\quad + \sum\limits_{ka} F_{ka} r^{\phantom{...}{a}}_{ijk} + \frac{1}{4} \sum\limits_{klcd} \langle kl || cd \rangle r^{\phantom{...}{cd}}_{ijkl}
\end{align*}

\begin{align*}
{\sigma}_{ijk}^{\phantom{...}{a}} &= \langle\Phi_{ijk}^{\phantom{...}{a}}|\bar{H}|R_1\Phi_{0}\rangle + \langle\Phi_{ijk}^{\phantom{...}{a}}|\bar{H}|R_2\Phi_{0}\rangle + \langle\Phi_{ijk}^{\phantom{...}{a}}|\bar{H}|R_3\Phi_{0}\rangle \\
&= P(k/ij) \left[ -\sum\limits_l F_{kl} r_{ijl}^{\phantom{...}{a}} + \frac{1}{2} \sum\limits_{lm} I^4_{ijlm} r_{lmk}^{\phantom{...}{a}} + \sum\limits_b h^{9r'}_{kb} t_{ij}^{ab} \right. \\
&\quad - \sum\limits_l I^2_{ijla} r_{kl}] - P(i/jk) \sum\limits_{bl} I^1_{ibla} r_{ljk}^{\phantom{...}{b}} + \sum\limits_b F_{ab} r_{ijk}^{\phantom{...}{b}} \\
&\quad + \sum\limits_{lb} F_{lb} r_{ijkl}^{\phantom{...}{ab}} + \frac{1}{2} \sum\limits_{lcb} I^7_{labc} r_{ijkl}^{\phantom{...}{cb}} - \frac{1}{2} P(k/ij)\sum\limits_{lmb} I^6_{lmkb} r_{ijlm}^{\phantom{...}{ab}} \\
&\quad + \frac{1}{2} \sum\limits_{dc} I^{5r}_{dc} t_{ijk}^{dca}
\end{align*}

\begin{align*}
{\sigma}_{ijkl}^{\phantom{...}{ab}} &= \langle\Phi_{ijkl}^{\phantom{...}{ab}}|\bar{H}|R_1\Phi_{0}\rangle 
+ \langle\Phi_{ijkl}^{\phantom{...}{ab}}|\bar{H}|R_2\Phi_{0}\rangle 
+ \langle\Phi_{ijkl}^{\phantom{...}{ab}}|\bar{H}|R_3\Phi_{0}\rangle \\
&= P(ab) \sum\limits_c F_{bc} r_{ijkl}^{\phantom{...}{ac}} 
- P(l/ijk) F_{lm} r_{ijkm}^{\phantom{...}{ab}} 
+ \frac{1}{2} \sum\limits_{cd} I^5_{abcd} r_{ijkl}^{\phantom{...}{cd}} \\
&\quad + \frac{1}{2} P(ij/kl) \sum\limits_{mn} I^4_{klmn} r_{ijmn}^{\phantom{...}{ab}} 
+ P(l/ijk) \sum\limits_c I^3_{lcba} r_{ijk}^{\phantom{...}{c}} \\
&\quad - P(ab) P(ij|kl) \sum\limits_{m} I^2_{klmb} r_{ijm}^{\phantom{...}{a}} 
- P(ab) P(l/ijk) \sum\limits_{mc} I^1_{lcma} r_{ijkm}^{\phantom{...}{cb}} \\
&\quad - P(l/ijk) \sum\limits_{m} H^{4r}_{ijkm} t_{ml}^{ab} 
+ P(ab) P(ij|kl) \sum\limits_{e} H^{5r}_{ijea} t_{kl}^{eb} \\
&\quad + P(i/jkl) \sum\limits_d H^{9r}_{id} t_{jkl}^{dab} 
- P(ik|jl) \sum\limits_{nd} H^{6r}_{idkn} t_{jnl}^{dab} \\
&\quad + \frac{1}{2} P(i/jkl) P(ab) \sum\limits_{cd} H^{7r}_{idca} t_{jkl}^{dcb}.
\end{align*}

Intermediates used for DIP:

\begin{align*}
I^{5r}_{cb} &= \frac{1}{2} \sum\limits_{ij} \langle ij || cb \rangle r_{ij} \\
H^{9r'}_{ia} &= \frac{1}{2} \sum\limits_{jk} I^6_{jkia} r_{jk} + \frac{1}{2} \sum\limits_{jkb} \langle jk || ba \rangle r_{jki}^{\phantom{...}{b}} \\
H^{9r}_{ia} &= \frac{1}{2} \sum\limits_{jk} I^6_{jkia} r_{jk} + \frac{1}{2} \sum\limits_{jkb} \langle jk || ba \rangle r_{jki}^{\phantom{...}{b}} - \sum\limits_{k} F_{ka} r_{ik} \\
H^{4r}_{jklm} &= P(l/jk)  \sum\limits_{ne}  I^6_{mnle} r_{jkn}^{\phantom{...}{e}} -  P(j/kl)\sum\limits_{n} I^4_{klnm} r_{jn} \\
& \quad + \frac{1}{2} \sum\limits_{nef} \langle nm || fe \rangle r_{jknl}^{\phantom{...}{fe}} \\
H^{5r}_{ijec} &= -P(ij) \sum\limits_{m} I^1_{iemc} r_{mj} - \frac{1}{2} P(ij) \sum\limits_{mn} I^6_{mnie} r_{jmn}^{\phantom{...}{c}} \\
& \quad + \sum\limits_{mf} I^7_{mcfe} r_{ijm}^{\phantom{...}{f}} - \frac{1}{2} \sum\limits_{mnf} \langle mn || ef \rangle r_{ijmn}^{\phantom{...}{cf}} \\
& \quad - \frac{1}{2} \sum\limits_{mnf} \langle nm || fe \rangle r_{mn} t_{ij}^{cf} \\
H^{6r}_{jdkn} &= -P(kj) \sum\limits_{m} I^6_{nmkd} r_{jm} - \sum\limits_{mf} \langle mn || df \rangle r_{jmk}^{\phantom{...}{f}} \\
H^{7r}_{idca} &= -\sum\limits_{m} I^7_{madc} r_{im} + \frac{1}{2} \sum\limits_{nm} \langle mn || cd \rangle r_{inm}^{\phantom{...}{a}} 
\end{align*}
For the intermediates one can refer to Levchenko et,al.\cite{levchenko2004equation} for( $F_{ij}$, $F_{ab}$, $F_{ia}$, $I^1$ $I^4$ $I^5$ $I^6$ and $I^7$  and  Manisha et,al.\cite{manohar2024spin} for ($I^2$ and $I^3$)

\end{document}